\documentclass{rmf-d}
\usepackage{nopageno,rmfbib,multicol,times,epsf,amsmath,amssymb,cite,braket}
\usepackage[T1]{fontenc} 
\usepackage[]{caption2}
\usepackage{graphicx}
\usepackage{blindtext}
\usepackage{color}
\usepackage{hyperref}
\usepackage{minted}
\DeclareRobustCommand{\openone}{\leavevmode\hbox{\small1\normalsize\kern-.33em1}}

\renewcommand{\Im}{\textrm{Im}}
\renewcommand{\Re}{\textrm{Re}}
\def\rmfcornisa{}

\def\rmfcintilla{}
\begin{document}

%
%
\title{The role of complementarity and quantum correlation in the uncertainty relations and the optimum states
\vspace{-6pt}}
\author{Sttiwuer D\'iaz-Sol\'orzano}
\address{Departamento de F\'isica, Universidad Sim\'on Bol\'ivar, Sartenejas, Estado Miranda 89000, Venezuela}
\author{Goosline Alfredo Pereira Narv\'aez}
\address{Departamento de F\'isica, Universidad Sim\'on Bol\'ivar, Sartenejas, Estado Miranda 89000, Venezuela}
%
%
\author{ }
\address{ }
\author{ }
\address{ }
\author{ }
\address{ }
\author{ }
\address{ }
\author{ }
\address{ }
\maketitle
%
%
\begin{abstract}
\textbf{Abstract.}
%
%
Uncertainty relations are used in quantum mechanics to illustrate the complementarity of two observables, assuming there is a lower bound for the product of their dispersions that depends on the first-order quantum correlation and the commutation relation of those observables in an optimum state. In this work, we demonstrate that the complex parameter of the optimum state equation has real and imaginary parts that depend on the first-order quantum correlation and the commutation relation between the observables in the optimum state, respectively. We apply our results to a two-level system, finding that the first-order quantum correlation and the commutation relation of two spin-1/2 projections in the optimum state complement each other and, finally, when the term correlation is used, a lower bound for the uncertainty principle does not depend on any parameter of the optimum state, in contrast to the usual form of the Heisenberg principle.
\vspace{1em}
\end{abstract}
\keys{ \bf{\textit{Quantum mechanics, Quantum correlation, Lower bound, Optimum states, Uncertainty relations
}} \vspace{-8pt}}
\begin{multicols}{2}

\section{Introduction}

Classically, it is always possible to design measurement devices for identically prepared systems such that the root-mean-square deviation (hereafter referred to as dispersion) of an observable's values can be reduced as much as desired, which is accomplished by fine-tuning the instruments to perform more precise measurements, or even by obtaining dispersion-free measurements. However, in quantum mechanics, the dispersions \footnote{We define $(\Delta \hat{\mathcal{A}})_{\psi}$ as the dispersion of an observable $\mathcal{\hat{A}}$ in the $\ket{\psi}$ state
\begin{equation*}
(\Delta \mathcal{\hat{A}})_{\psi}=\sqrt{\braket{\mathcal{\hat{A}}^2}_{\psi}-\braket{\mathcal{\hat{A}}}_{\psi}^2}.
\end{equation*}} of two incompatible observables cannot simultaneously approach zero when measured in identical systems, due to the complementarity and the quantum correlation between these observables. Von Neumann \cite{von-neumann}
posed the question: Is there a relation between the dispersions of incompatible observables that prevents them from simultaneously tending to zero? In 1927, Heisenberg \cite{heisenberg} answered this question mathematically for canonically complementary observables (e.g., position and linear momentum), finding a lower bound for the product of dispersions independent of the state $\ket{\psi}$,
\begin{equation}\label{eq:0.3}
(\Delta\hat{\mathcal{A}})^2_{\psi}(\Delta\hat{\mathcal{B}})^2_{\psi}\ge\frac{\hbar^2}{4},
\end{equation}
here $\hat{\mathcal{A}}$ and $\hat{\mathcal{B}}$ are two canonically conjugated observables such that $[\hat{\mathcal{A}},\hat{\mathcal{B}}]=\textrm{i} \hbar\hat{\openone}$. However, this relation does not explicitly show the complementarity between the two observables at the lower bound. In 1929, Robertson~\cite{robertson} generalized this relation to two observables that may or may not be canonically conjugated (i.e., $[\hat{\mathcal{A}},\hat{\mathcal{B}}]\neq\textrm{i}\hbar\hat{\openone}$), exhibiting a complementarity term at a lower bound, associated with the commutation relation between the two observables. The case when this relation is fulfilled by the observables shows a statistically independent distributions (i.e., uncorrelated), where the Robertson relation is given by,
\begin{equation}\label{eq:0.4}
(\Delta\hat{\mathcal{A}})^2_{\psi}(\Delta\hat{\mathcal{B}})^2_{\psi}\geq%
\left|\frac{1}{2}\braket{\textrm{i}[\hat{\mathcal{A}},\hat{\mathcal{B}}]}_{\psi}\right|^2.
\end{equation}
Here, a lower bound for the dispersion product of two observables depends on the pure state. Subsequently, works from Schr\"odinger~\cite{Schrodinger1930}, Gamba~\cite{gamba1950}, and Castoldi~\cite{castoldi}
found that a lower bound given by Robertson relation is increased by a term associated with the anticommutator of
these observables, which is
\begin{eqnarray}\label{eq:0.1}
\begin{aligned}
(\Delta\hat{\mathcal{A}})^2_{\psi}(\Delta\hat{\mathcal{B}})^2_{\psi}%
&\geq%
\left|\frac{1}{2}\braket{\textrm{i}[\hat{\mathcal{A}},\hat{\mathcal{B}}]}_{\psi}\right|^2
\\&+
\left|\frac{1}{2}\braket{\{\hat{\mathcal{A}},\hat{\mathcal{B}}\}}_{\psi}-%
\braket{\hat{\mathcal{A}}}_{\psi}\braket{\hat{\mathcal{B}}}_{\psi}\right|^2.
\end{aligned}
\end{eqnarray}
The second term corresponds to correlations between the statistical distributions of the values thrown by the observables after their measurement, and these correlations are obtained from the quantum correlation \cite{Bohm-libro} or Pearson correlation coefficient \footnote{The Pearson correlation coefficient, $C_{\mathcal{A}\mathcal{B}}$, is

\begin{equation*}
C_{\mathcal{A}\mathcal{B}}=\frac{\braket{\mathcal{A}\mathcal{B}}-\braket{\mathcal{A}}\braket{\mathcal{B}}}{(\Delta\mathcal{A})(\Delta\mathcal{B})},
\end{equation*}
where $(\Delta\mathcal{A})$ and $(\Delta\mathcal{B})$ are the variances of the classical observables $\mathcal{A}$ and $\mathcal{B}$, respectively. On the other hand, the non-normalized Pearson's correlation coefficient is given by

\begin{equation*}
C_{\mathcal{A}\mathcal{B}}=\braket{\mathcal{A}\mathcal{B}}-\braket{\mathcal{A}}\braket{\mathcal{B}}.
\end{equation*}} (non-normalized). It is worth mentioning that before Gamba and Castoldi, Schr\"odinger had a particular interest in this new term (correlation) in the cases of operators with continuous spectrum. This correlation measures the degree of joint variation in the distribution of values of two observables and the linear relation between the distribution of those values.
\\%

\par\indent%
The characterization of the uncertainty principle as a limitation of quantum preparations
and measurements has led to a widespread view that claims this principle is nothing but a
formal expression of the principle of complementarity \cite{Busch.Heinonen.Lahti.2007}.
This allows expression $\frac{1}{2}\braket{\textrm{i}[\hat{\mathcal{A}},\hat{\mathcal{B}}]}_{\psi}$ to be interpreted as a complementarity term. Thus, the quantum covariance and the complementarity term present a degree of competition at a lower bound of the uncertainty relation.
That is, both quantities have a trade-off to generate the Heisenberg relation \eqref{eq:0.3} even when the observables are not complementary and are not canonically conjugated. This contrasts with the established literature \cite{klauder-skagerstam, sakurai-napolitano, Bohm-libro, gerry-knight, merzbacher, cohen-diu-laloe, peres}, where the lower bound for the product of the dispersions of two observables is obtained when they are uncorrelated and \eqref{eq:0.4} is saturated. However, when there is a quantum correlation between the observables, the lower bound increases and is obtained by saturating \eqref{eq:0.1}. That is what happens in the case of spin-1/2 projections 
in the optimum states \cite{optimumstates} or minimum uncertainty states,  for which a lower bound of the uncertainty relation is obtained. This work presents a theoretical development for the construction of optimum states, and we find these states for two arbitrary spin-1/2 projections.
\\%

\par\indent%
This paper is organized as follows: in Sec. II, we briefly review the correlation between two observables called quantum correlation. In Sec. III, we review optimum states and derive the complex parameter of the optimum state equation for any pair of observables. In Sec. IV, we analyze the arbitrary spin-1/2 projections and show that there is no difference between a lower bounds of the Schrödinger-Gamba-Castoldi uncertainty relations and that of Heisenberg when the system is prepared in its optimum states. Finally, in Sec. V, we present our conclusions.

\section{Quantum correlation}
The correlation (covariance) between two classic observables $\mathcal{A}$ and $\mathcal{B}$ is given by
\begin{equation}\label{eq:1.1}
\textrm{Cov}(\mathcal{A},\mathcal{B})=%
\braket{\mathcal{AB}}-\braket{\mathcal{A}}\braket{\mathcal{B}},%
\end{equation}
where $\braket{\bullet}$ represent the classic expected value. Through the canonical quantization rule, the product of classic observables is promoted to the semi-sum of the symmetric product of both quantum operators, i.e., $\mathcal{AB}\to\frac{1}{2}(\hat{\mathcal{A}}\hat{\mathcal{B}}+\hat{\mathcal{B}}\hat{\mathcal{A}})\equiv\frac{1}{2}\{\hat{\mathcal{A}},\hat{\mathcal{B}}\}$  and the classic mean values are replaced by quantum mean values. Thus, the correspondence principle is implicit at the limit $\hbar\to0$, where the classical observables commute and their quantum mean values approximates the corresponding classical expected values. However, the principle of correspondence is not rigorous or complete since there are observables that do not have classic analogous~\cite{Yndurain-libro}. In this sense, the first-order quantum correlation between two quantum observables $\mathcal{\hat{A}}$ and $\mathcal{\hat{B}}$ in the pure state $\ket{\psi}$ is~\cite{Schrodinger1930,Bohm-libro}
\begin{equation}\label{eq:1.2}
\textrm{C}_{1,1}(\hat{\mathcal{A}},\hat{\mathcal{B}};\psi)=%
\frac{1}{2}\braket{\{\hat{\mathcal{A}},\hat{\mathcal{B}}\}}_\psi-%
\braket{\hat{\mathcal{A}}}_\psi\braket{\hat{\mathcal{B}}}_\psi,
\end{equation}
where the subscripts (1,1) indicate that the correlation is to first order. Substituting \eqref{eq:1.2} in \eqref{eq:0.1}, we have that the Schr\"odinger-Gamba-Castoldi uncertainty relation takes the following form
\begin{equation}\label{eq:1.4}
(\Delta\hat{\mathcal{A}})^2_{\psi}(\Delta\hat{\mathcal{B}})^2_{\psi}%
\geq%
\left|\frac{1}{2}\braket{\textrm{i}[\hat{\mathcal{A}},\hat{\mathcal{B}}]}_{\psi}\right|^2+%
\left|\textrm{C}_{1,1}(\hat{\mathcal{A}},\hat{\mathcal{B}};\psi)\right|^2.
\end{equation}
The first expression of \eqref{eq:1.4} will be called the complementarity relation, being two complementary observables if the conmutation relation is a c-number, i.e., $[\hat{\mathcal{A}},\hat{\mathcal{B}}]=c_0\hat{\openone}$.
\\%

\par\indent%
On the other hand, when two observables are statistically independent, the first-order quantum correlation $\textrm{C}_{1,1}(\hat{\mathcal{A}},\hat{\mathcal{B}};\psi)$ is zero. However, this does not indicate that no further correlations exist. In order to measure all types of correlations between two observables in the same quantum system, Bohm~\cite{Bohm-libro} sets out to investigate the cancellation of the expression:
\begin{equation}\label{eq:1.3}
\textrm{C}_{m,n}(\hat{\mathcal{A}},\hat{\mathcal{B}};\psi)=%
\frac{1}{2}\braket{\{\hat{\mathcal{A}}^m,\hat{\mathcal{B}}^n\}}_\psi-%
\braket{\hat{\mathcal{A}}^m}_\psi\braket{\hat{\mathcal{B}}^n}_\psi,%
\end{equation}
for all $m,n\in\mathbb{N}=\{1,2,\ldots\}$. It is important to note that if the state $\ket{\psi}$ is an eigenvector of either of the two observables, the quantum correlation \eqref{eq:1.3} vanishes, such that in this preparation, both observables are uncorrelated. 
\section{Review of Optimum states}
The developments in references \cite{gamba1950} and \cite{castoldi} show that optimum states are those that saturate the Cauchy-Schwarz inequality for the scalar product between vectors in the Hilbert space $\mathcal{H}$. A quantum state $\ket{\psi_\textrm{op}}$ will be called optimum state of two observables $\hat{\mathcal{A}}$ and $\hat{\mathcal{B}}$, if the state belongs to the domain of both observables and its image is proportional when the observables act on the state $\ket{\psi_{\textrm{op}}}$
\begin{equation}\label{eq:2.1}
\hat{\mathcal{A}}\ket{\psi_\textrm{op}}=\lambda\hat{\mathcal{B}}\ket{\psi_\textrm{op}},%
\end{equation}
with $\lambda\in\mathbb{C}$. It is important to note that the optimum state equation \eqref{eq:2.1} does not suggest that the observables be considered proportional. In general, the constant of proportionality $\lambda$ is a non-zero complex number, which can be obtained from the mean values of the commutation and anticommutation relations of the o\-pe\-ra\-tors involved. In fact, using \eqref{eq:2.1} and its adjoint
(i.e., $\bra{\psi_\textrm{op}}\hat{\mathcal{A}}=\bar{\lambda}\bra{\psi_\textrm{op}}\hat{\mathcal{B}}$)
we can obtain,
\begin{subequations} \label{eq:2.2}
\begin{eqnarray}
\label{eq:2.2a}
&&\braket{\{\hat{\mathcal{A}},\hat{\mathcal{B}}\}}_{\psi_\textrm{op}}=%
2\Re[\lambda]\braket{\hat{\mathcal{B}}^2}_{\psi_\textrm{op}},
\\%
\label{eq:2.2b}
&&\braket{\textrm{i}[\hat{\mathcal{A}},\hat{\mathcal{B}}]}_{\psi_\textrm{op}}=%
2\Im[\lambda]\braket{\hat{\mathcal{B}}^2}_{\psi_\textrm{op}}.
\end{eqnarray}
\end{subequations}

\noindent The real and imaginary parts of the parameter $ \lambda$, shown in \eqref{eq:2.1}, are written in terms of the correlation \eqref{eq:1.2} and the mean value of the commutator of these ob\-ser\-va\-bles, respectively. Indeed, using \eqref{eq:1.2}, \eqref{eq:2.1} and \eqref{eq:2.2} we can write $ \lambda$
\begin{eqnarray}\label{eq:2.3}
\begin{aligned}
\lambda&=
\frac{\textrm{C}_{1,1}(\hat{\mathcal{A}},\hat{\mathcal{B}};\psi_{\textrm{op}})}{(\Delta\hat{\mathcal{B}})_{\psi_\textrm{op}}^2}+%
\textrm{i}\frac{\braket{\textrm{i}[\hat{\mathcal{A}},\hat{\mathcal{B}}]}_{\psi_\textrm{op}}}{2(\Delta\hat{\mathcal{B}})_{\psi_\textrm{op}}^2}.
\end{aligned}
\end{eqnarray}
From the equality presented in \eqref{eq:2.3}, it is clear that $ \lambda$ is pure imaginary when the two observables are uncorrelated to first order. This fact is consistent with \eqref{eq:0.4}. However, if $ \lambda \in \mathbb{R}^*$ then 
$\braket{\textrm{i}[\hat{\mathcal{A}},\hat{\mathcal{B}}]}_{\psi_{\textrm{op}}}=0$. In this case, \eqref{eq:0.4} does not impose a limit for the product of the dispersions, despite the fact that the observables exhibit first-order quantum correlation. On the other hand, \eqref{eq:1.4} imposes a bound for the product of these dispersions given by \eqref{eq:1.2}.
\\%

\par\indent%
Now, let us evaluate the optimum state at a lower bound for the product of the square of the dispersions of the observables $ \hat{\mathcal{A}}$ and $ \hat{\mathcal{B}}$. To do so, we substitute \eqref{eq:2.1} and \eqref{eq:2.2} into \eqref{eq:0.1}
\begin{eqnarray}\label{eq:2.4}
\begin{aligned}
(\Delta\hat{\mathcal{A}})_{\psi_\textrm{op}}^2(\Delta\hat{\mathcal{B}})_{\psi_\textrm{op}}^2%
&=%
\left|\lambda(\Delta\hat{\mathcal{B}})_{\psi_\textrm{op}}^2+\frac{\bar{\lambda}-\lambda}{2}\braket{\hat{\mathcal{B}}^2}_{\psi_\textrm{op}}\right|^2\\&+%
\left|\frac{\bar{\lambda}-\lambda}{2}\right|^2\braket{\hat{\mathcal{B}}^2}_{\psi_\textrm{op}}^2.
\end{aligned}
\end{eqnarray}
From the result presented in \eqref{eq:2.4}, it is possible to conclude: (I) When the constant $\lambda$ is a real number (i.e., $\lambda=\lambda_0$ with $ \lambda_0\in\mathbb{R}^*$), then the product of the dispersions coincides with the quantum correlation of both observables in the optimum state,
\begin{equation}\label{eq:2.5}
(\Delta\hat{\mathcal{A}})_{\psi_{\textrm{op}}}(\Delta\hat{\mathcal{B}})_{\psi_{\textrm{op}}}=%
\left|\textrm{C}_{1,1}(\hat{\mathcal{A}},\hat{\mathcal{B}};\psi_{\textrm{op}})\right|.
\end{equation}
In other words, from \eqref{eq:2.4} we obtain that $|\lambda_0|=(\Delta\hat{\mathcal{A}})_{\psi_{\textrm{op}}}/(\Delta\hat{\mathcal{B}})_{\psi_{\textrm{op}}}$ and, substituting $ \lambda=\lambda_0$ in \eqref{eq:2.3} the expression \eqref{eq:2.5} is reached.
\\%

\par\indent%
(II) When $ \lambda$ is a pure imaginary number (i.e., $\lambda=\textrm{i} \lambda_0$ with $\lambda_0 \in \mathbb{R}^*$), then the expression \eqref{eq:2.4} takes the following form:
\begin{equation*}
2\lambda_0^2\braket{\hat{\mathcal{B}}^2}_{\psi_{\textrm{op}}}%
\left[(\Delta\hat{\mathcal{B}})_{\psi_{\textrm{op}}}^2-\braket{\hat{\mathcal{B}}^2}_{\psi_{\textrm{op}}}\right]=0,
\end{equation*}

\noindent where $(\Delta\hat{\mathcal{A}})_{\psi_{\textrm{op}}}=|\lambda_0|(\Delta\hat{\mathcal{B}})_{\psi_{\textrm{op}}}$ has been used. Taking into account that the solution $\braket{\hat{\mathcal{B}}^2}_{\psi_{\textrm{op}}}\neq0$, since this  leads to a negative dispersion, then
\begin{equation}
\label{eq:2.6}
(\Delta\hat{\mathcal{B}})_{\psi_{\textrm{op}}}^2=\braket{\hat{\mathcal{B}}^2}_{\psi_{\textrm{op}}}
\;\therefore\;%
\braket{\hat{\mathcal{B}}}_{\psi_{\textrm{op}}}=\braket{\hat{\mathcal{A}}}_{\psi_{\textrm{op}}}=0.
\end{equation}

So the only possible solution is that the observables involved have zero mean values, when $ \lambda$ is a pure imaginary number. In general, the reciprocal is not true: if the mean values of the observables are zero, it does not imply that $ \lambda$ is a pure imaginary number.
\\%

\par\indent%
Finally, to determine the existence of the optimum states, it is sufficient to verify whether the kernel of the operator $\hat{\mathcal{A}}- \lambda \hat{\mathcal{B}}$ is empty or not. If it is empty, then there are no non-trivial states that verify equation \eqref{eq:2.1}, otherwise, the dimension of the kernel will account for the number of optimum states associated with the observables involved. In addition, if the observable $\hat{\mathcal{B}}$ has an inverse, then the optimum states correspond to an eigenvector of $ \hat{\mathcal{B}}^{-1} \hat{\mathcal{A}}$ with eigenvalue $ \lambda$. If the observable $\hat{\mathcal{A}}$ has an inverse, then the optimum states $\ket{\psi_{\textrm{op}}}$ corresponds to an eigenvector of $\hat{\mathcal{A}}^{-1} \hat{\mathcal{B}}$ with eigenvalue $\lambda^{-1}$. 

\subsection*{Optimum states for a two-level system}
Consider two self-adjoint operators $ \hat{\mathcal{A}}$ and $ \hat{\mathcal{B}}$ defined on a two-dimensional Hilbert space $ \mathcal{H}_2$ where the observable $ \hat{\mathcal{B}}$ is non-singular (i.e., $ \det(\hat{\mathcal{B}})\neq0$), thus verifying the existence of its inverse and the fact that none of the values that the observable acquires after its measurement are zero. Given this hypothesis, we can apply $ \hat{\mathcal{B}}^{-1}$ to \eqref{eq:2.1} such that the optimum states of the observables $\hat{\mathcal{A}}$ and $ \hat{\mathcal{B}}$ correspond to the eigenvectors of the operator $ \hat{\mathcal{B}}^{-1}\hat{\mathcal{A}}$. Now, to obtain the eigenvalues we have to impose the condition $\det{(\hat{\mathcal{B}}^{-1}{\hat{\mathcal{A}}}-\lambda{\hat{\openone}})}=0$. Then the characteristic polynomial takes the following form
\begin{equation}
\label{eq:59}
\lambda^2-\textrm{tr}(\hat{\mathcal{B}}^{-1}\hat{\mathcal{A}})\lambda+\det(\hat{\mathcal{B}}^{-1}\hat{\mathcal{A}})=0.
\end{equation}

Solving the quadratic equation for the $\lambda$ presented in \eqref{eq:59}, we obtain
\begin{equation}\label{eq:60}
\lambda_{\pm}=\frac{\textrm{tr}(\hat{\mathcal{B}}^{-1}\hat{\mathcal{A}})
\pm\sqrt{\left[\textrm{tr}(\hat{\mathcal{B}}^{-1}\hat{\mathcal{A}})\right]^2
-4\det(\hat{\mathcal{A}}\hat{\mathcal{B}})}}{2\det(\hat{\mathcal{B}})}.
\end{equation}

Let us consider that the optimum state can be wri\-tten in the canonical basis $ \{\ket{i}\} \in \{\ket{0}, \ket{1}\}$ as $ \ket{\psi_{\textrm{op}}^{\pm}}=\sum_{i}Z_i^{\pm}\ket{i}$, where the sign $+$ ($-$) corresponds to $ \lambda_+$ ($ \lambda_-$). To determine it, it is sufficient to define the singular (non-invertible) operator
\begin{equation}\label{eq:65}
\hat{\mathcal{N}}_{\pm}\equiv%
\left[\textrm{tr}(\hat{\mathcal{B}})\hat{\openone}-\hat{\mathcal{B}}\right]\hat{\mathcal{A}}
-\det(\hat{\mathcal{B}})\lambda_\pm\hat{\openone},
\end{equation}
where $ \lambda_{\pm}$ is given in \eqref{eq:60} and $N^{\pm}_{ij}=\braket{i|\hat{\mathcal{N}}_{\pm}|j}$ co\-rres\-ponds to the matrix elements of \eqref{eq:65} in the canonical basis. Then the components of the optimum state take the following form
\begin{equation}\label{eq:66}
\sum_{j=0}^1 N^{\pm}_{ij}Z_j^\pm=0%
\;\Rightarrow\;%
\left\{\begin{array}{ll}
Z_0^\pm=N_{01}^{\pm}=N_{11}^\pm
\\%
Z_1^\pm=-N_{00}^{\pm}=-N_{10}^\pm
\end{array}\right. .
\end{equation}

Therefore, the optimum states are given by
\begin{equation}\label{eq:1.7}
\ket{\psi_\textrm{op}^\pm}=%
\frac{N^{\pm}_{01}\ket{0}-N^{\pm}_{00}\ket{1}}{\sqrt{|N^{\pm}_{00}|^2+|N^{\pm}_{01}|^2}}.%
\end{equation}

 \section{Optimum states: spin-1/2 systems}
A particle’s spin-1/2 is described by the vector operator $ \vec{\mathbb{S}}=\hat{\mathbb{S}}_1\vec{\imath}+\hat{\mathbb{S}}_2\vec{\jmath}+\hat{\mathbb{S}}_3\vec{k}$, where the set of unit vectors $ \{\vec{\imath}, \vec{\jmath}, \vec{k}\}$ is the canonical basis of the Euclidean space $ \mathbb{R}^3$, while the spin-$ 1/2$ projections in these directions are given by
\begin{eqnarray}
\begin{aligned}
\label{eq:4.1}
&\hat{\mathbb{S}}_1=\frac{\hbar}{2}\hat{\sigma}_1=\frac{\hbar}{2}%
\left(\begin{array}{cc}
0 & 1 \\
1 & 0 \\
\end{array}\right),
\\&
\hat{\mathbb{S}}_2=\frac{\hbar}{2}\hat{\sigma}_2=\frac{\hbar}{2}%
\left(\begin{array}{cc}
0 &-\textrm{i} \\
\textrm{i} & 0 \\
\end{array}\right),
\\&
\hat{\mathbb{S}}_3=\frac{\hbar}{2}\hat{\sigma}_3=\frac{\hbar}{2}%
\left(\begin{array}{cc}
1&0 \\
0&-1\\
\end{array}\right),
\end{aligned}
\end{eqnarray}

\noindent where $ \hat{\sigma}_i$ ( $ i\in \{1,2,3\}$) are the Pauli matrices. The spin-$ 1/2$ projection along the direction of the unit vector $  \vec{m}= \sin(\theta_m)\cos(\varphi_m)\vec{\imath}+\sin(\theta_m)\sin(\varphi_m)\vec{\jmath}+\cos(\theta_m)\vec{k}$ is given by the expression
\begin{eqnarray}\label{eq:4.2}
\begin{aligned}
\hat{\mathbb{S}}_m&=\frac{\hbar}{2}\vec{m}\cdot\vec{\sigma}=%
\frac{\hbar}{2}\sum_{j=1}^3m_j\hat{\sigma}_j%
\\&=\frac{\hbar}{2} \left(\begin{array}{cc}
  \cos(\theta_m) & \textrm{e}^{-\textrm{i}\varphi_m}\sin(\theta_m) \\
 \textrm{e}^{\textrm{i}\varphi_m}\sin(\theta_m) & -\cos(\theta_m) 
\end{array}\right).
\end{aligned}
\end{eqnarray}

\noindent Here, the angles $(\theta_m,\varphi_m)$ are the spherical coordinates of the unit vector $ \vec{m}$. It is clear that \eqref{eq:4.2} verifies all the expressions given in \eqref{eq:4.1} when the unit vector $ \vec{m}$ is chosen among the directions $ \vec{\imath}$, $ \vec{\jmath}$ or $ \vec{k}$, respectively. Considering that
\begin{equation}
\label{eq:4.6}
(\vec{a}\cdot\vec{\sigma})(\vec{b}\cdot\vec{\sigma})=
\vec{a}\cdot\vec{b}\,\hat{\openone}+\textrm{i}(\vec{a}\times\vec{b})\cdot\vec{\sigma},
\end{equation}
the square of the operator \eqref{eq:4.2} is given by
\begin{equation}\label{eq:1.6}
\hat{\mathbb{S}}_m^2=\frac{\hbar^2}{4}(\vec{m}\cdot\vec{\sigma})(\vec{m}\cdot\vec{\sigma})=%
\frac{\hbar^2}{4}\hat{\openone},
\end{equation}
and the commutator and anticommutator for two arbitrary spin-$ 1/2$ projections are given by
\begin{subequations} \label{eq:4.4}
\begin{eqnarray}
\label{eq:4.4a}
\left[\hat{\mathbb{S}}_m,\hat{\mathbb{S}}_n\right]&=&%
\textrm{i}\hbar||\vec{m}\times\vec{n}||\,\hat{\mathbb{S}}_{m\times n},
\\%
\label{eq:4.4b}
\left\{\hat{\mathbb{S}}_m,\hat{\mathbb{S}}_n\right\}&=&%
\frac{\hbar^2}{2}\vec{m}\cdot\vec{n}\,\hat{\openone}.
\end{eqnarray}
\end{subequations}

\noindent In addition, the eigenvalues of \eqref{eq:4.2} are $ \pm\frac{\hbar}{2}$ and the eigenvectors are given by
\begin{subequations} \label{eq:4.3}
\begin{eqnarray}
\label{eq:4.3a}
\hspace{-.5cm}\ket{\psi_{m}^{+}}&=&%
\cos\left(\frac{\theta_m}{2}\right)\ket{0}+\textrm{e}^{\textrm{i}\varphi_m}\sin\left(\frac{\theta_m}{2}\right)\ket{1},%
\\%
\label{eq:4.3b}
\hspace{-.5cm}\ket{\psi_{m}^{-}}&=&%
-\sin\left(\frac{\theta_m}{2}\right)\ket{0}+\textrm{e}^{\textrm{i}\varphi_m}\cos\left(\frac{\theta_m}{2}\right)\ket{1},%
\end{eqnarray}
\end{subequations}

\noindent where $ \ket{0}$ and $ \ket{1}$ are the spin-$ 1/2$ states along the $ z$ direction. The ket $ \ket{\psi_m^{+}}$ represents the state of a particle whose spin is polarized or parallel to the $ \vec{m}$ direction (spin up), while $ \ket{\psi_m^{-}}$ is the state with opposite orientation (spin down). The states $ \ket{\psi_m^{+}}$ and $ \ket{\psi_m^{-}}$ are or\-tho\-go\-nal states, such that $ \ket{\psi_m^{-}}$ is obtained from $ \ket{\psi_m^{+}}$ by the transformation $ \theta_m\to\theta_m+\pi$. Thus, $ \vec{m}$ and $ -\vec{m}$ are the Fano-Bloch vectors for the pure states $ \ket{\psi_m^{+}}$ and $ \ket{\psi_m^{-}}$, respectively. The Fano-Bloch vector of a pure state is obtained by means of the expression \cite{fano.1957}
\begin{equation}
\label{eq:4.5}
\vec{n}_\psi=\braket{\vec{\sigma}}_{\psi}=\braket{\psi|\vec{\sigma}|\psi}.
\end{equation}
If $\ket{\psi}=\ket{\psi^{\pm}_m}$ then $\vec{n}_\psi=\pm\vec{m}$.

\subsection*{Optimum state for two spin-$ 1/2$ projections}
To evaluate the optimum states corresponding to two arbitrary spin-$ 1/2$ projections we will consider in \eqref{eq:2.1} that $ \hat{\mathcal{A}}=\hat{\mathbb{S}}_m$ and $\hat{\mathcal{B}}=\hat{\mathbb{S}}_n$. Then, the following operator is required to obtain the parameter  $ \lambda$ and the optimum state
\begin{eqnarray}\label{eq:108}
\begin{aligned}
\hat{\mathcal{B}}^{-1}\hat{\mathcal{A}}&=%
\hat{\mathbb{S}}_n^{-1}\hat{\mathbb{S}}_m=%
(\vec{n}\cdot\vec{\sigma})(\vec{m}\cdot\vec{\sigma})
\\&
=(\vec{n}\cdot\vec{m})\,\hat{\openone}+\textrm{i}(\vec{n}\times\vec{m})\cdot\vec{\sigma}.
\end{aligned}
\end{eqnarray}

\noindent The optimum states are obtained by the eigenvectors $ \ket{\psi^{\pm}_{\textrm{op}}}$ of the \eqref{eq:108} operator, where $ \ket{\psi^{+}_{\textrm{op}}}$ and $ \ket{\psi^{-}_{\textrm{op}}}$ are oriented along the polarization vectors $ \vec{n}_{\textrm{op}}$ and $ -\vec{n}_{\textrm{op}}$, respectively. The polarization vector is given by
\begin{equation}\label{eq:1.5}
\vec{n}_{\textrm{op}}\equiv%
\frac{\vec{n}\times\vec{m}}{||\vec{n}\times\vec{m}||}=%
-\frac{\vec{m}\times\vec{n}}{||\vec{m}\times\vec{n}||}.
\end{equation}
By finding the eigenvalues of the operator \eqref{eq:108}, we obtain the parameter $ \lambda_\pm$ of the optimum state equation
\begin{equation}\label{eq:3.17}
\lambda_\pm=\vec{m}\cdot\vec{n}\pm\textrm{i}||\vec{m}\times\vec{n}||.
\end{equation}
This result is in complete agreement with \eqref{eq:60}. On the other hand, the operator $ \hat{\mathcal{N}}_{\pm}$, defined in \eqref{eq:65}, is proportional to the projector constructed with the optimum state, that is
\begin{equation}\label{eq:1.8}
\hat{\mathcal{N}}_\pm=%
\pm\textrm{i}\frac{\hbar^2}{4}||\vec{m}\times\vec{n}||\ket{\psi_{\textrm{op}}^{\mp}}\bra{\psi_{\textrm{op}}^{\mp}}.
\end{equation}
Moreover, the real part of \eqref{eq:3.17} contains information on the correlation between observables, while its imaginary part contains information on the mean value of the co\-mmu\-ta\-tor. Then, from the equations \eqref{eq:2.2} and the mean value of \eqref{eq:1.6} we get
\begin{subequations} \label{eq:2.10}
\begin{eqnarray}
\label{eq:2.10a}
\braket{\textrm{i}[\hat{\mathbb{S}}_m,\hat{\mathbb{S}}_n]}_{\psi_{\textrm{op}}^{\pm}}&=&%
\pm\frac{\hbar^2}{2}||\vec{m}\times\vec{n}||,%
\\%
\label{eq:2.10b}
\braket{\{\hat{\mathbb{S}}_m,\hat{\mathbb{S}}_n\}}_{\psi_{\textrm{op}}^{\pm}}&=&%
\frac{\hbar^2}{2}\left(\vec{m}\cdot\vec{n}\right).
\end{eqnarray}
\end{subequations}

These quantities complement each other, since
\begin{equation}\label{eq:3.18}
(\vec{m}\cdot\vec{n})^2+||\vec{m}\times\vec{n}||^2=1.
\end{equation}

\subsection*{A lower bound for dispersion of two spin-1/2 projections}
The mean value of the observable \eqref{eq:4.2} in the pure state $ \ket{\psi}$, with Fano-Bloch vector $ \vec{n}_\psi$, is given by
\begin{equation}\label{eq:100}
\braket{\hat{\mathbb{S}}_m}_\psi=%
\frac{\hbar}{2}\braket{\vec{m}\cdot\vec{\sigma}}_\psi=%
\frac{\hbar}{2}\,\vec{m}\cdot\vec{n}_\psi%
=\frac{\hbar}{2}\,\vec{m}\cdot(\vec{n}_\psi-\vec{n}_{\textrm{op}}),
\end{equation}

\noindent where the identity \eqref{eq:4.5} and the fact that $ \vec{m}\cdot\vec{n}_{\textrm{op}}=0$ have been used; thus, the mean value  \eqref{eq:100} is zero in the optimum state. When the $ \ket{\psi}$ state is polarized in the spin direction $ \ket{\psi_m^{\pm}}$, i.e., $ \vec{n}_{\psi}=\pm\vec{m}$, we have $ \braket{\hat{\mathbb{S}}_m}_{\psi}=\pm\frac{\hbar}{2}$. Using \eqref{eq:100}, the mean value in the $\ket{\psi}$ state of \eqref{eq:4.4} gives the result
\begin{subequations} \label{eq:105}
\begin{eqnarray}
\label{eq:105a}
\hspace{-.5cm}%
\braket{\textrm{i}[\hat{\mathbb{S}}_m,\hat{\mathbb{S}}_n]}_\psi&=&%
\frac{\hbar^2}{2}||\vec{m}\times\vec{n}||\,\vec{n}_{\textrm{op}}\cdot\vec{n}_\psi,
\\%
\label{eq:105b}
\hspace{-.5cm}%
\braket{\{\hat{\mathbb{S}}_m,\hat{\mathbb{S}}_n\}}_\psi&=&%
\frac{\hbar^2}{2}\vec{m}\cdot\vec{n}.
\end{eqnarray}
\end{subequations}

\noindent It is clear that the mean value of the anticommutator is independent of the state $ \ket{\psi}$, while the mean value of the commutator depends on the polarization state $ \ket{\psi}$. These quantities coincide with \eqref{eq:2.10} when $ \ket{\psi}=\ket{\psi^{\pm}_{\textrm{op}}}$. With the quantities \eqref{eq:100} and \eqref{eq:105} it is possible to evaluate the quantum correlation \eqref{eq:1.2} of two spin-$ 1/2$ projections in arbitrary directions, according to
\begin{eqnarray}
\begin{aligned}
\label{eq:106}
&\textrm{C}_{1,1}(\hat{\mathbb{S}}_m,\hat{\mathbb{S}}_n;\psi)=%
\frac{\hbar^2}{4}\,(\vec{m}\times\vec{n}_\psi)\cdot(\vec{n}\times\vec{n}_\psi)
\\&
=\frac{\hbar^2}{4}\left[\vec{m}\cdot\vec{n}
-\vec{m}\cdot\left(\vec{n}_\psi-\vec{n}_{\textrm{op}}\right)%
\vec{n}\cdot\left(\vec{n}_\psi-\vec{n}_{\textrm{op}}\right)\right]^2,
\end{aligned}
\end{eqnarray}

\noindent where $\vec{m}\cdot\vec{n}_{\textrm{op}}=\vec{n}\cdot\vec{n}_{\textrm{op}}=0$. The covariance \eqref{eq:106} is zero if $ \vec{m}$ is in the plane formed by the vectors $ \vec{n}_{\psi}$ and $ \vec{n}\times{\vec{n}}_{\psi}$. A particular case in which the first-order quantum correlation is zero occurs when the three polarization vectors $ \vec{m}$, $ \vec{n}$ and $ \vec{n}_{\psi}$ are orthogonal. For example, it occurs when the Fano-Bloch vectors $ \vec{m}$, $ \vec{n}$ and $ \vec{n}_{\psi}$ correspond in the observables $ \hat{\mathbb{S}}_1$, $ \hat{\mathbb{S}}_2$ and $ \hat{\mathbb{S}}_3$, respectively. While the maximum value of the correlation \eqref{eq:106} occurs when $ \vec{n}_{\psi}$ reaches the optimum state \eqref{eq:1.5}.
\\%

\par\indent%
On the other hand, using \eqref{eq:4.2} and \eqref{eq:4.6} we can evaluate the dispersion of the spin-$ 1/2$ projection in the $ \ket{\psi}$ state
\begin{equation}\label{eq:103}
(\Delta\hat{\mathbb{S}}_m)_\psi=%
\sqrt{\braket{\hat{\mathbb{S}}_m^2}_\psi-\braket{\hat{\mathbb{S}}_m}_\psi^2}=%
\frac{\hbar}{2}||\vec{m}\times\vec{n}_\psi||.
\end{equation}

The dispersion vanishes when $ \ket{\psi}$ is polarized along the spin direction (i.e., when $ \vec{n}_{\psi}=\pm\vec{m})$. Now, with the expression \eqref{eq:103}, one can evaluate the product of the dispersions of two spin-$ 1/2$ projections in either direction $ \vec{m}$ and $ \vec{n}$, resulting in
\begin{equation}\label{eq:110}
(\Delta\hat{\mathbb{S}}_m)_\psi^2(\Delta\hat{\mathbb{S}}_n)_\psi^2=\frac{\hbar^4}{16}%
||\vec{m}\times\vec{n}_\psi||^2||\vec{n}\times\vec{n}_\psi||^2.%
\end{equation}

While a lower bound of the product of the dispersions given by \eqref{eq:1.4} is
\begin{eqnarray}\label{eq:111}
\begin{aligned}
&\left|\textrm{C}_{1,1}(\hat{\mathbb{S}}_m,\hat{\mathbb{S}}_n;\psi)\right|^2+%
\left|\frac{1}{2}\braket{\textrm{i}[\hat{\mathbb{S}}_m,\hat{\mathbb{S}}_n]}_{\psi}\right|^2 \\&=%
\frac{\hbar^4}{16}\left[\left((\vec{m}\times\vec{n}_\psi)\cdot(\vec{n}\times\vec{n}_\psi)\right)^2+\left(\vec{m}\cdot(\vec{n}_\psi\times\vec{n})\right)^2\right].
\end{aligned}
\end{eqnarray}

In general, the expressions \eqref{eq:110} and \eqref{eq:111} are not the same, they only coincide when the optimum state \eqref{eq:1.5} is reached (i.e., when $ \vec{n}_{\psi}=\vec{n}_{\textrm{op}}$). Then, a lower bound for the product of dispersions of two spin-$ 1/2$ projections takes the form of the Heisenberg relation \eqref{eq:0.3}. In the optimum state, the complementary relation ($ \braket{\textrm{i}[\hat{\mathbb{S}}_m,\hat{\mathbb{S}}_n]}_{\psi_{\textrm{op}}}$) and the first-order quantum covariance $ \textrm{C}_{1,1}(\hat{\mathbb{S}}_m,\hat{\mathbb{S}}_n; \psi_{\textrm{op}})$ complement each other, since the equation \eqref{eq:3.18} is reached. Therefore,
\begin{equation}\label{eq:2.9}
(\Delta\hat{\mathbb{S}}_m)^2_{\psi_{\textrm{op}}}(\Delta\hat{\mathbb{S}}_n)^2_{\psi_{\textrm{op}}}=%
\frac{\hbar^2}{4}.
\end{equation}

\section{Conclusions}
A lower bound in Heisenberg’s uncertainty relation \eqref{eq:0.3} is preparation-independent on which the measurement of the observables is performed, showing its invalidity when the preparation is an eigenstate of one of the two observables, since it leads to the inconsistency $ 0 \geq\frac{\hbar^2}{4}$. This suggests that a lower bound is not independent of the preparation. Such a difficulty does not show up in the uncertainty relations \eqref{eq:0.4} and \eqref{eq:0.1}. If the preparation is an eigenstate of one or both of the two observables, Robertson's uncertainty relation \eqref{eq:0.4} has no difficulty, verifying equality. Similarly, the Schrödinger-Gamba-Castoldi uncertainty relation also presents no difficulty, since the first-order quantum correlation between the observables is zero, as $ \braket{\{\hat{\mathcal{A}},\hat{\mathcal{B}}\}}_{\psi}=2\braket{\hat{\mathcal{A}}}_{\psi}\braket{\hat{\mathcal{B}}}_{\psi}$. In this case, the observables are free of dispersion and reproduce classical properties, i.e., the dispersion that the observables may have is due to an imperfection in the measuring apparatus and can be brought to zero; that means that measurements of the observables in compatible preparations do not disturb the measured value. 
\\%

\par\indent%
On the other hand, by excluding the possibility that $ \ket{\psi}$ is a proper state of the observables involved, the relations \eqref{eq:0.3}, \eqref{eq:0.4} and \eqref{eq:0.1} show that the product of the dispersion of the observables cannot tend to zero simultaneously even if the measurement is as accurate as possible. This is due to the quantum nature of the observables and the correlation that may exist between them, not to the imperfection with which they have been measured; that is, the measurement of the observables in such preparations perturbs the measured value. In fact, when there is correlation between two compatible observables (they commute), relation \eqref{eq:0.1} avoids the product of the dispersions to be zero. In conclusion, the uncertainty relations \eqref{eq:0.3}, \eqref{eq:0.4} and \eqref{eq:0.1} mathematically formalize the uncertainty principle or the complementarity principle: \emph{Two incompatible and correlated observables cannot be measured simultaneously in the same experiment without the product of their dispersions being as far from zero as po\-ssi\-ble}. The advantage of the Schrödinger-Gamba-Castoldi uncertainty relation \eqref{eq:0.1} over the Robertson uncertainty relation \eqref{eq:0.4} is that the former considers the first-order quantum correlation between observables. Furthermore, the Robertson relation is incompatible with the optimum state equation unless the first-order quantum correlation between the observables vanishes. For the case of a spin-$ 1/2$ system, it is observed that the uncertainty relation \eqref{eq:0.1} coincides with \eqref{eq:0.3} in the optimum state, this does not occur if we consider the relation \eqref{eq:0.4} because the spin-$ 1/2$ projections exhibit first-order quantum correlation.
\\%

\par\indent%
To investigate a lower bound of the product of dispersions of two observables, it is necessary to know the optimum state. These states verify the equality in the Cauchy-Schwarz relation for the scalar product defined in the Hilbert space of the physical system under con\-si\-de\-ra\-tion. The constant $ \lambda$ of the optimum state equation \eqref{eq:2.1} carries with it the information of the commutation relation and the first-order quantum correlation of the observables in their imaginary and real parts, respectively. This leads to the fact that $\lambda$ is real if the commutation relation between the observables in the optimum state is zero, on the contrary, it will be pure imaginary when the first-order quantum correlation between those observables is zero in the optimum state. However, when the mean va\-lues of the observables are zero in the optimum state, it does not necessarily imply that $\lambda$ is pure imaginary.

\section{Appendix}
To prove the expression \eqref{eq:2.3} we will use the fact that
\begin{equation} \label{eq:ap1}
\braket{\hat{\mathcal{A}}}_{\psi_{\textrm{op}}}=\lambda\braket{\hat{\mathcal{B}}}_{\psi_{\textrm{op}}}
\end{equation}
under the optimum state equation \eqref{eq:2.1}. In addition, we will use the expressions \eqref{eq:1.2} and \eqref{eq:2.2}, then
\begin{eqnarray*} 
&&\lambda=\Re[\lambda]+\textrm{i}\Im[\lambda]=\frac{1}{2}\frac{\braket{\{\hat{\mathcal{A}},\hat{\mathcal{B}}\}}_{\psi_{\textrm{op}}}}{\braket{\hat{\mathcal{B}}^2}_{\psi_{\textrm{op}}}}+\textrm{i}\frac{1}{2}\frac{\braket{\textrm{i}[\hat{\mathcal{A}},\hat{\mathcal{B}}]}_{\psi_{\textrm{op}}}}{\braket{\hat{\mathcal{B}}^2}_{\psi_{\textrm{op}}}}
\\
\Rightarrow 
&&\braket{\hat{\mathcal{B}}^2}_{\psi_{\textrm{op}}}\lambda=
\frac{1}{2}\braket{\{\hat{\mathcal{A}},\hat{\mathcal{B}}\}}_{\psi_{\textrm{op}}}+\textrm{i}\frac{1}{2}\braket{\textrm{i}[\hat{\mathcal{A}},\hat{\mathcal{B}}]}_{\psi_{\textrm{op}}}
\\
&&=\textrm{C}_{1,1}(\hat{\mathcal{A}},\hat{\mathcal{B}};\psi_{\textrm{op}})+\braket{\hat{\mathcal{A}}}_{\psi_{\textrm{op}}}\braket{\hat{\mathcal{B}}}_{\psi_{\textrm{op}}}+\textrm{i}\frac{1}{2}\braket{\textrm{i}[\hat{\mathcal{A}},\hat{\mathcal{B}}]}_{\psi_{\textrm{op}}}
\\
&&=\textrm{C}_{1,1}(\hat{\mathcal{A}},\hat{\mathcal{B}};\psi_{\textrm{op}})+\lambda\braket{\hat{\mathcal{B}}}_{\psi_{\textrm{op}}}^2+\textrm{i}\frac{1}{2}\braket{\textrm{i}[\hat{\mathcal{A}},\hat{\mathcal{B}}]}_{\psi_{\textrm{op}}}
\\ \Rightarrow  &&\lambda(\braket{\hat{\mathcal{B}}^2}_{\psi_{\textrm{op}}}-\braket{\hat{\mathcal{B}}}_{\psi_{\textrm{op}}}^2)=\textrm{C}_{1,1}(\hat{\mathcal{A}},\hat{\mathcal{B}};\psi_{\textrm{op}})+\textrm{i}\frac{1}{2}\braket{\textrm{i}[\hat{\mathcal{A}},\hat{\mathcal{B}}]}_{\psi_{\textrm{op}}}
\\
\Rightarrow
&&\lambda(\Delta \hat{\mathcal{B}})^2_{\psi_{\textrm{op}}}=\textrm{C}_{1,1}(\hat{\mathcal{A}},\hat{\mathcal{B}};\psi_{\textrm{op}})+\textrm{i}\frac{1}{2}\braket{\textrm{i}[\hat{\mathcal{A}},\hat{\mathcal{B}}]}_{\psi_{\textrm{op}}}
\\
\therefore &&
\lambda=\frac{\textrm{C}_{1,1}(\hat{\mathcal{A}},\hat{\mathcal{B}};\psi_{\textrm{op}})}{(\Delta \hat{\mathcal{B}})^2_{\psi_{\textrm{op}}}}+\textrm{i}\frac{\braket{\textrm{i}[\hat{\mathcal{A}},\hat{\mathcal{B}}]}_{\psi_{\textrm{op}}}}{2(\Delta \hat{\mathcal{B}})^2_{\psi_{\textrm{op}}}}.
\end{eqnarray*}
\vspace{1mm}

\section*{Acknowledgments}

We would like to dedicate this article to the memory of Luis A. Gonz\'alez-D\'iaz, friend and physicist.

\end{multicols}

\medline
\begin{multicols}{2}
%
\nocite{*}
\bibliographystyle{rmf-style}
\bibliography{ref}
%
%
%

%
%
\end{multicols}
\end{document}